\documentclass{article}
\pdfoutput=1                

\usepackage[utf8]{inputenc} 
\usepackage[T1]{fontenc}    
\usepackage{hyperref}       
\usepackage{url}            
\usepackage{booktabs}       
\usepackage{amsfonts}       
\usepackage{nicefrac}       
\usepackage{microtype}      
\usepackage{graphicx}       
\usepackage{authblk}        
\usepackage{parskip}
\usepackage{float}

\usepackage[verbose=true,letterpaper]{geometry}
\AtBeginDocument{
  \newgeometry{
    textheight=9in,
    textwidth=6.5in,
    top=1in,
    headheight=14pt,
    headsep=25pt,
    footskip=30pt
  }
}

\title{Radical Realism}

\author{Noah Guzmán}
\affil{Independent Scholar}

\author{Nicolás Hinrichs}
\affil{FrameNet Brasil Computational Linguistics Lab., Federal University of Juiz de Fora, Brazil}
\affil{Institute of Applied Linguistics and Translatology, Leipzig University, Germany}
\affil{Psycholinguistics Laboratory, University of Concepción, Chile}
\affil{nicolashinrichs@protonmail.ch}

\date{January 24, 2024}

\begin{document}
\maketitle

\begin{abstract}
The ontogeny of cognitive neuroscience has emerged within the hegemony of substance ontology. Persistent physicalist influences are described through three developmental hallmarks that yielded epistemic attractors - promoters and perpetuators of material-discursive practices oriented toward reification and self-vindication across the interdisciplinary spectrum which, as a whole, has been driven away from its pretensions to scientific realism. In virtue of a desire for a radical return thereto, we adopt a metaphysic stance akin to pragmatism, and briefly make the case that such concerns have sociopolitical implications extending far beyond the realm of mere philosophical interest.
\end{abstract}

\section{Introduction}\label{sec1}


Emil du Bois-Reymond, co-discoverer of the action potential, referred to Darwin as the “Copernicus of the organic world” \cite{du_bois-reymond_darwin_1883} during his address delivered at the anniversary meeting of the Berlin Academy of Sciences. Freud was inspired by the comparison \cite{finkelstein_emil_2013} and is attributed to have proclaimed the three “insults to humanity”, which he pinpointed as stemming from heliocentrism, evolution, and psychoanalysis. The existential realisations we take closer to heart, are those stemming from du Bois’ earlier proposition, by thinking less of them as affronts triggering a narcissistic hurt, but as hallmarks of humanities’ glimpses of that which is unamendable (we will aim to elucidate unamendability as the core concept for cognitive neuroscience to move closer towards its pretensions to realism).

In similar fashion as du Bois did, we propose three further moments, as they set the initial conditions within which cognitive neuroscience emerged and currently develops in; moreover, we highlight them alongside their meta-theoretical manifestation, conforming with our notion of epistemic attrators, which act as main promoters of said ideas: firstly, physicalism as proto-objectivity, alongside substance ontology. Secondly, reification by means of the divorce between ethology and behaviourism and, thirdly, the true psychic turn by the hand of the cognitive revolution and surge of Machine Learning and Artificial Intelligence. 

The first two events will be explored in the second section, while the third event will be addressed in section four under the rubric of \textit{neuroliberalism}, as will be the resistance against it - which a radical realism can offer. Section three highlights our notion of epistemic attractors. Section five closes this manuscript on sense-making practices, pragmatist realism and the mind-independence of success.

\section{The Ontogeny of Neuroscientific Metaphysics}\label{sec2}

\begin{quote}
  "As far as the laws of mathematics refer to reality, they are not certain; and as far as they are certain, they do not refer to reality." \cite{Einstein2005-EINGAE}
\end{quote}

\subsection{The Phantom of Physicalism}\label{subsec1}


The ontogeny of the cognitive sciences has unfolded atop a scaffolding collectively referred to as \textit{substance metaphysics}, which refers to a diverse set of views that understand reality as consisting of different unchanging objects or entities, each composed of more fundamental objects, often with some objects serving as a ``most fundamental''  base-level of reality. Though substance perspectives have implicitly composed a metaphysical hegemony since the ancient Greek philosophers, it is not until the 20th century when the presumption of stability became a dominant  position scientists sought to conform with - as did the explanation of change as mere alterations in the configuration of inherently stable substances -as opposed to change being regarded as fundamental, with any ``thing-like'' entities corresponding to steady-states of processes-, which came hand in hand with several shifts in epistemic virtues (including the likes of control, calculation, consensus, context, collaboration, critical cross-checking of sources, etc.). 

The mechanisms by which these regularized are not only scarcely expressed as explicit rules but also often purposefully refrained from being addressed at all, which has kept a wide range of disciplines unaware of the fact that it is the epistemic scaffoldings -out of which scientific paradigms co-emerge, in relation to human experience and, more importantly so, to its apparent limits- which give epistemology its metaphorical character \footnote{As implied by Foucault \cite{foucault_order_1971}: ``The episteme is not a form of knowledge (connaissance) or type of rationality which, crossing the boundaries of the most varied sciences, manifests the sovereign unity of a subject, a spirit, or a period; it is the totality of relations that can be discovered, for a given period, between the sciences when one analyses them at the level of discursive regularities.''}. It follows that research programs \cite{lakatos_methodology_1999} are guided by axioms, analogies, metonyms, etc. - without which we would not be able to scientifically analyse phenomena. This has become particularly evident in the cognitive sciences, as definitions, visualisations, and data collection for the development of models of cognition have had to employ several constructs in order to deduct what is not directly observable. Hacking \cite{hacking_representing_1983} goes as far as pointing out that

\begin{quote}
``Experimentation has a life of its own, interacting with speculation, calculation, model building, invention and technology in numerous ways.''
\end{quote}

Yet rarely do practitioners of neuroscience express awareness of their implicit subscription to any particular metaphysical position, despite being brought up in a tradition with its own language shaped by a long history of theorizing, which influences their activities by biasing them towards certain directions; implicit metaphysical assumptions are ``baked-in'' to the metaphors and other linguistic tools scientists use to reason about the world and guide their activities in it. In the words of Oderberg \cite{oderberg_real_2007},

\begin{quote}
    ``Natural language is permeated and saturated by metaphysics, and has been so ever since philosophy began with the pre-Socratics... The problem is in thinking that there is a vantage point from which we can espy language in its ‘ordinary’, pre-metaphysical state. There is no such vantage point because there is no such language to be observed in the first place.''
\end{quote}

Indeed, epistemic virtues such as positivist truth-to-nature and physicalist objectivity hide, ignore, and/or even actively refuse to acknowledge their own naturalistic foundational issues stemming from an implicit acceptance of substance metaphysical assumptions, an attitude that has trickled down into- and plagued the contemporary cognitive and behavioral neurosciences, manifesting as a neurocentric essentialism\footnote{That is, the behavior of an organism situated in its environment is reified as a set of neural functions carried out by specific brain regions or networks, with any actual interaction between the organism and its environment seen as an uninteresting black box to be solved by those working on motor control and sensory physiology} and an exaggerated emphasis on neurocomputational explanations of behavior, by which functional processes internal to the organism constitute a reification of processes distributed across the brain, body, and environment \cite{chemero_radical_2013, di_paolo_sensorimotor_2017, newen_embodied_2018}, which cannot be localized to the organism alone, thereby erecting a barrier to the study of behavior by allowing investigators to simply write promissory notes in place of missing explanations rather than elucidating the holistic nature of the mechanisms giving rise to such behavior.

The cognitive sciences have indeed adhered for a long time now to a widespread epistemic ideal of sufficiency in interdisciplinarity understood as an interaction between its disciplines that lead to accounts of phenomena  at different levels of observation -and realism- with the purpose of providing a complete explanation of cognition. This principle guided both neurocentrism and neurocomputationalism since the cognitive revolution's programme, but not only have, for instance, Marr's levels of analysis\footnote{i.e., the computational, which maps in/output of tasks; algorithmic, which defines the rules for this; and the implementational, which describes the material instantiation of both.}\cite{marr_vision_2010} proven to be difficult to accommodate most frameworks \cite{harkness_moving_2017} yet also the attempt to cover absolute explanations of cognitive phenomena in all of them seems to forcibly forfeit actual integration of conceptual models with non-coterminous theories \cite{andersen_collaboration_2016}, thus reinforcing their axiomatic coherence by operational closure of discourse practice\footnote{As Kording and colleagues \cite{kording_appreciating_2020} point out: “Models positioned at different levels of biological realism (microscopic, macroscopic, behavioral, representational) are not guaranteed to inform each other, as distinct phenomena may emerge at different levels”.}. 

\subsection{Divorce and Reification}\label{subsec2}

The divorce between ethology and behaviourism since the mid 20th century brought about reification as its main mechanism of progress. As noted by Fagot-Largeault \cite{berthoz_anthropological_2009}, in the anglophone world, studies of behavior had been pursued primarily by psychologists employing mouse or rat model organisms in tightly controlled laboratory settings using artificially constructed discrete tasks (the behaviorists), whereas in the non-anglophone world, the study of behavior had typically been undertaken by zoologists observing the continuous behavior of a diverse array of species in their natural ecological contexts (the ethologists). The split between these two schools of thought only grew in the era following the second world war. The styles of reasoning employed by behaviorism and its later successor, cognitive psychology, lent themselves to the systematic reification of behavior; in seeking for experimental control in the laboratory, practitioners generated a large number of artificial behavioral phenomena novel to the laboratory setting and assumed the existence of a hypothetical set of mental functions which might be responsible for these phenomena. The burgeoning fields of cognitive and behavioral neuroscience took many of their methodological and stylistic cues from behaviorism and cognitive psychology and designed a number of partially fictional mechanisms for these functions, mechanisms for which they sought to manufacture evidence. 

Since we have accused the neurosciences of reification, it is necessary to provide at least a sketch of what we mean by the term. The parameters which enable such closure stem from adherence to physicalism and its main resource to enable the recurring of such an epistemic mistake is reification. Mel Andrews writes \cite{andrews_making_2022}

\begin{quote}
    "(...) the mismapping of formal structure onto target phenomena—or theoretical representation thereof—in a manner that leads us to misapprehend the causal structure of nature. Reification is an epistemic error. It involves, by definition, a fallacious inference."
\end{quote}

Them calling the mapping of formal structure to reality an epistemic mistake is correct and this error indeed is commonplace, and can be formalised to shed light on the broader malady we subscribe it emerges within as a wrongful praxis pervasive to most scientific practices. Further, in this broad sense of inference-based cross-domain mappings, such an error is often obscured under the term "metaphor"; Sfard \cite{Sfard1994ReificationAT} even refers to reification as the necessary birth of metaphor. Metaphors metaphorised in this way are bound to dualistic contingency, as in doing so one \cite{16/3/265.diaz}

\begin{quote}
"hides the circumstances of the emergence of idiosyncratic metaphorizing in (...) subjects, i.e., the characterization of the conditions that foster, if not trigger, hinder, or even thwart this emergence."
\end{quote}

A commonplace central argument of scientific realism (expressed as a discipline-based variant of the broader philosophical discussion, which we’ll refer to in subsequent sections) states that the properties of a model are expected to correspond to an ontological property of the phenomenon of interest, might this take form as literal presence of properties of the model in its target systems \cite{Chakravartty2017-CHARON-4, rescorla_realist_2019} or some isomorphism in the representation \cite{ramstead_is_2020} which stands in as sign of its pursuance of an ultimate explanatory truth about the phenomena. It is worth to note at this point that naturalistic programs such as minimal physicalism \cite{fields_minimal_2021} will embody Maddy’s idea of a \cite{maddy_second_2007, maddy_defending_2011} \textit{Second Philosopher}, insofar as their empirical inquiries will be anointed an exclusive albeit full operational closure by the mathematical domain; as it delivers a mechanism of verification for said examinations with disregard for any particular \textit{liaison} (or \textit{cul-de-sac}, for that matter) between sources of ontology and epistemology, these remain neglected as “the special sciences”.

Hence, we'll further examine the particular sense of reification stemming from marxist canon on ideology that is frequently omitted in its contemporary usage; namely, it was originally proposed as a special kind of alienation. If one were to even further stress the case that there is indeed a prevalent form of fallacious argumentation in discursive practices pertaining to the neuroscientific production niches, then we still ought to examine how to conciliate this reformulation as an improper praxis which results from a cognocentrist bias which runs through the entirety of Western metaphysics, as already suggested by the likes of Heidegger and Dewey. Lukács \cite{lukacs_history_1971} provides a definition which closer resembles Marx's in referring to reification as 

\begin{quote}
``a relation between persons [that] takes on the character of a thing and thus acquires a 'phantom objectivity,' a seemingly rational autonomy which borders with autarky in its concealment of all traces of its core constituent, namely, the relation between persons.''
\end{quote}

In other words, reification is the process by which social or subjective qualities are mistakenly perceived as objective or independent objects. It follows reification cannot be summed up solely as a strictly epistemic or scientific mistake \cite{honneth_reification_2008}:

\begin{quote}
``This is not only because reification constitutes a multilayered and stable syndrome of distorted consciousness, but also because this shift in attitude reaches far too deep into our habits and modes of behavior for it to be able to be reversed by making a corresponding cognitive correction.'' 
\end{quote}

This attentiveness (or loss thereof), seen as an antecedent act to cognition, can be understood as a kind of "empathetic engagement" or, in current day theoretical-neuroscientific discourse, perhaps akin to Pauen's \textit{perspectivalness} \cite{pauen_second-person_2012}, that is, the ability which allows us to take over a second person's perspective, as one both chronologically and ontologically prior to our ability to cognize them at all. It follows that reification emerges within a belief system which scaffolds into detachment, solipsism, and objectification of relationships as mere data-driven transactions; this was conflated by the twilight of self-critical judgments of reason, coming about during the second half of the twentieth century by the hand of rule-bound reason\footnote{In fact, it is precisely human interiority which was castrated from the cognitive agenda, as it didn't quite fit any particular granularity the program deemed proper to address, as its orienting ``informational'' metaphor was based on a substance principle that permeated the newly proposed social practices, if we understand these as ``embodied skills that have a common style and are transposed to various domains'' and ``have a unity and form a social field'' \cite{dreyfus_rainbow_bourdieu}. Themes, in this sense, can be understood as embodied, \textit{a priori} assumptions acting as scaffolding templates for the construction of other cognitive descriptors. Following this, the particular issue of interiority can be further explained by the ``Relations between inner and outer'' \cite{keane_semiotic_2018}, of which one could argue that they are problems only as they come to be thematised in ``specific semiotic ideologies, given certain political, religious, or other historically specific circumstances.''}; moreover, by the advent of algorithmic rules of rationality \cite{daston_rules_2022}. 

Algorithmic metaphysics were carried unto neuroscience by means of the prior historical emergence of scientific symbolic notation wherein the discipline was born, expressing itself as an epistemic bias in the form of a commitment towards this 'unattentive' mode of reifying behavioral phenomena, and ultimately yielding ontic naturalisms unable to properly account for non-physical properties\footnote{As Avanessian and Hennig \cite{avanessian_metanoia_2018} have put it: ``(...) what language makes visible is not so much the existence of things but the existence of relations.''}.  An example of this semiotic disavowal of realism in cognitive neuroscience that has received notoriety is \textit{encodingism} \cite{brette_is_2019}. Leibniz noted that the permanence of symbolic writings -which nothing in effective mathematics \textit{a priori} requests- governs contingent methodological procedures as it constitutes a primordial requirement of symbolic notation that transcends the requisites of any immediate signification. Its projection as an epistemic guide for research has been called the scheme of invariance-extension or \textit{principe de prolongement} \cite{serfati_methode_2002} and as a symbolic language it is ontologically distinct from any natural language; yet, one can proceed with is as in a formal, uninterpreted calculation, a possibility that grants this procedure its success and is guaranteed by the establishment, from the beginning, of a correspondence between characters and ‘things’. This can be construed as the continuation of the Newtonian tradition of reductionism in taking a structure as abstraction of the parts of the system and executing a dynamic model through it to infer its properties. It is precisely the matter of structural reductionism where realism in cognitive neuroscience finds it crux. 

We are now in a position to elucidate the central crisis of neuroscience (in truth, the central crisis of all of the sciences of behavior): the practices of reification defining the neurosciences are decidedly anti-realist, undermining the field's realist pretensions. Investigators have lost sight of the field’s original goal of explaining the behavior of organisms in their natural contexts. They first observe some behavioral phenomenon and then wish to investigate its neural underpinnings. But as they think of controlled ways to study the phenomenon, they are inevitably coming up with half-baked theories about its neural underpinnings. These half-baked theories become new, reinterpreted descriptions of the phenomenon that are different than the phenomenon itself. Investigators then end up studying phenomena that have nothing to do with the original one of interest because they end up designing experiments to study those imagined functions or mechanisms that they believe instantiate or give rise to the original phenomenon; such experiments, rather than reproducing the original phenomenon, produce new and often entirely unrelated phenomena which become the focus of study. These novel phenomena produced in the laboratory can only be connected back to the original phenomena by some vague homomorphic mapping (which we might call “task isomorphism”) postulated by the investigators. Not only do such investigations fail to explain the original real patterns of behavior first observed, they lead to theories whose style of reasoning obeys an internal logic that cannot be verified with respect to external domains of philosophical and scientific practices. Faye \cite{Faye2022-FAYNBE} warns us that the ascription of epistemic properties entering in complementary pairs is of course, only meaningful relative to an experimental set-up; moreover, their ascription to an object as existing independently of that specific experimental interaction should be regarded as ill-defined, a warning which becomes even more pressing to be taken under consideration at the intersection of scientific realism and instrumentalism.

\section{Epistemic Attractors}\label{sec3}
Having given a historical perspective on the metaphysical foundations of neuroscience, our task now is to analyze how the neurosciences have evolved toward their current ontoepistemological crisis. Some might wonder what right have we to use the word ``crisis'' here. Where is the evidence of this calamity? Though the eruptions of Mt. Vesuvius and the fall of the Roman empire were both predicted by soothsayers, only the volcanic inferno was seen as a catastrophe in its own time, while the inhabitants of the empire continued to feast and applaud themselves during its slow decline. The crisis in neuroscience has more in common with the rot of the Roman empire than the demolition of Pompeii. 

We can analogize the progression of the neurosciences, and really any science or set of practices for interacting with the world in general, as a dynamical system evolving on a manifold in what we will refer to as an \textit{epistemic space}. In this metaphor, it is useful to view our previous account of the fateful origins of neuroscience in a flawed substance metaphysics as providing a set of initial conditions for the modern evolution of the neurosciences within epistemic space. Though the notion of a mathematical space has been used primarily metaphorically in philosophy (see e.g. Sellars' ``space of reasons'' \cite{sellars_space_2007}), we hope to bring the idea of an epistemic space closer to the literal by providing a sketch of how such a space might be quantified by observable variables. 

To define an epistemic space, we will use Hasok Chang's tripartite activity-based analysis of science \cite{chang_realism_2022} which decomposes actions in the real of scientific investigation into \textit{operations}, \textit{epistemic activities}, and \textit{systems of practice}. At the moment, we will only define these ostensively before returning to them later in a more formal context. Operations are generally motor movements such as moving an organism under a microscope, pulling glass electrodes, connecting a measuring device to a computer, or typing on a keyboard to program a data analysis routine. We often string many of these actions together into complete activities, such as running an fMRI experiment or interpreting the results of a study. Each of these day-to-day, often repeated, activities can be collected together at the scale of a single scientist or an entire field of science to define a system of practice. Notably, though we are here interested in science, these concepts are not restricted to scientific domains but can be used to characterize systems of practice in philosophy or even daily life. In this analysis, we see the coordinates of epistemic space as being observable aspects or components of operations; at the most granular level, these variables could even be the kinematics of the bodily movements involved in some operation such as pipetting a solution.  

\begin{figure}[H]
    \centering
    \includegraphics[scale=0.25]{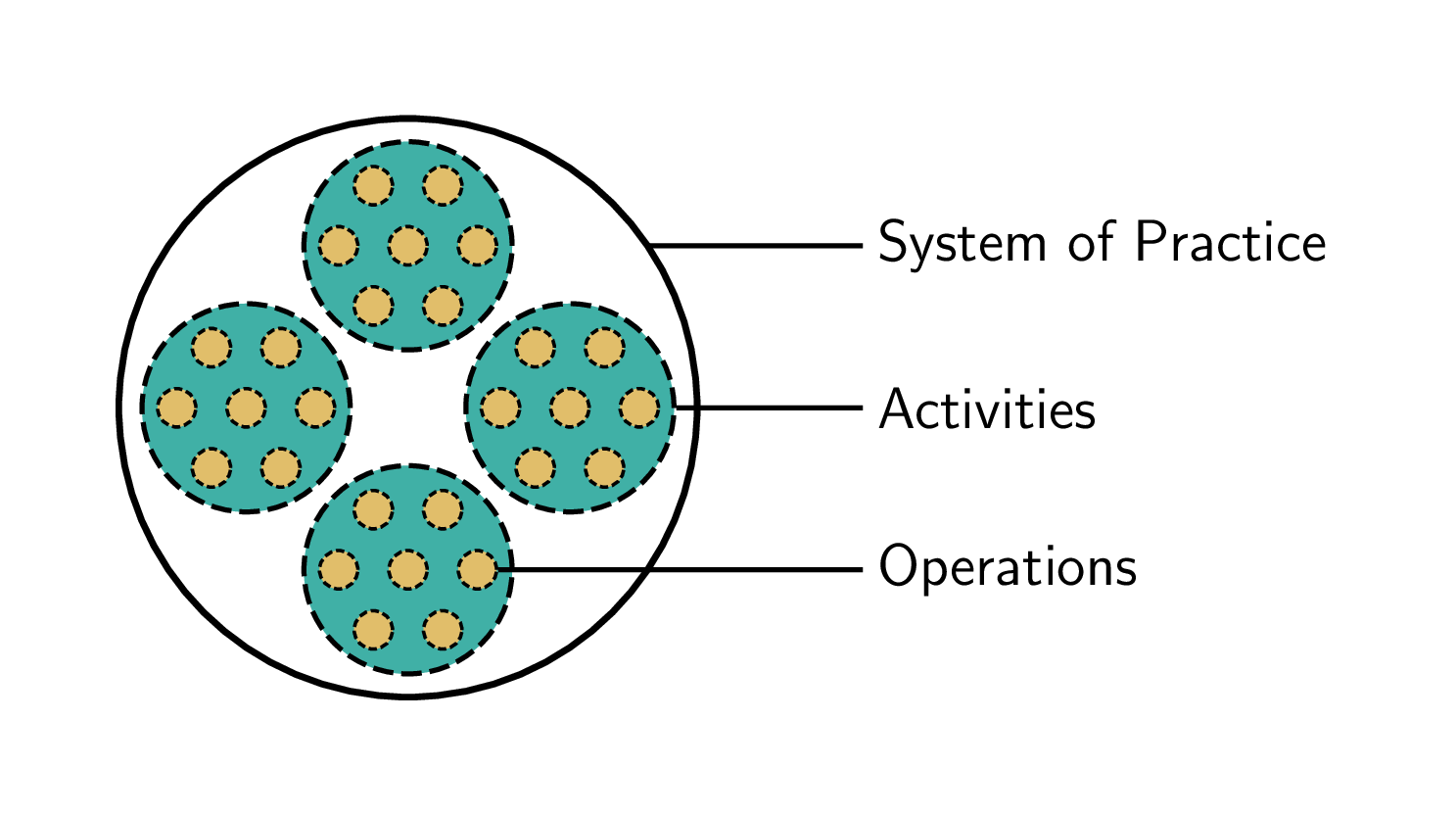}
    \caption{Schematic representation of Hasok Chang's three-level activity-based analysis of science. Adapted from Soler and Catinaud \cite{soler:hal-01878099}.}
    \label{fig:sop}
\end{figure}

We begin our exploration of the ``laws'' governing motion on epistemic manifolds with the Duhem-Quine thesis, also known as confirmation holism. The Duhem-Quine thesis is a philosophical view about the nature of scientific theories and their relationship to evidence. It was first proposed by Pierre Duhem in the context of physics and later developed and extended to the special sciences and other domains of investigation by Willard Van Orman Quine. According to the Duhem-Quine thesis, a scientific theory cannot be confirmed or refuted by deriving and testing predictions from a single hypothesis, only the whole bundle of hypotheses and assumptions that make up the scientific framework within which the theory is embedded. This means that, based on how one accommodates the experimental results, there could be multiple theories which make incompatible claims about the world but which are equally well supported by the data. Hypotheses in the bundle which are separate from the few hypotheses usually tested in an experiment are called \textit{auxiliary hypotheses}.

Philosophers of science have traditionally responded to Duhem and Quines' concerns about holism in science by examining the ways in which auxiliary hypotheses may be adjusted to account for a mismatch between experiment and theory. But during the ``practice turn'' in the philosophy of science, philosophers began analyzing the ways in which the day-to-day material practices of scientists reflect alternative strategies for responding to empirical evidence that contradicts theoretical predictions. Concerned with analyzing the cultural dynamics of science and their relation to confirmation holism, Ian Hacking's account of the processes of \textit{self-vindication} in the laboratory sciences is a story about scientists' diverse responses to contradictory experimental evidence \cite{hacking_2_2010}.

Self-vindication is discussed by Hacking as an emergent phenomenon of the ``laboratory style of thinking''. It will be helpful here to discuss what Hacking means by a ``style of thinking'' or ``style of reasoning''\cite{hacking_language_1982, hacking_style_1992}. This task is more difficult than it may initially seem since Hacking himself, as well as the historian Crombie from whom the concept is derived, only demonstrate styles via specific examples, relying on the family resemblance among these styles of reasoning to implicitly define what a style is. We think of styles of reasoning in terms of Gibson's notion of affordances \cite{gibson_senses_1983}; systems of practice are characterized by the affordances, possibilities for action specified by ecological information, which are tied to their activities and operations. The idea of a style of reasoning captures the ways in which linguistic conventions become coupled to affordances for thinking as well as more bodily actions, specifically those actions involved in the verification of some speech act. Here we mean verification in the positivist sense of ascertaining the truth of a proposition via interaction with the world. As certain modes of speech acts (styles of reasoning) develop and become coupled to affordances, new propositions become candidates for truth or falsehood since before that style of reasoning, no methods existed for their verification. As an example, consider an individual who had never been introduced to the ``postulational'' style of reasoning exemplified by early Greek geometry; until this person had been trained in the methods and ways of thinking common to ancient geometers, the request to ``Prove that triangle $A$ and triangle $B$ are similar'' would be completely meaningless to them because they were not yet familiar with the linguistic conventions coupling such a request to the affordances involved in deducing conclusions from geometric axioms.

Styles of reasoning \cite{hacking_representing_1983}, regarded as commitments towards tacit knowledge based discovery, are referred to as explanatory styles by contemporary pluralist accounts \cite{potochnik_patterns_2020}, having already been addressed in similar fashion as thought styles \cite{fleck_genesis_2008}, epistemes \cite{foucault_archaeology_2010}, paradigms \cite{kuhn_structure_2012, popper_myth_1997} and research programmes \cite{lakatos_methodology_1999, feyerabend_philosophical_1981}. We proffer that any such notion should be best complemented by analogy of Marcel Mauss’s critique against the idea of cultural diffusion \cite{levi-strauss_introduction_1987}, namely, that it is based on the assumption that movement of people, technologies, and ideas is somehow unusual. Instead, the exact opposite is true: there are entire realms of exchange being navigated at intervals of diverse regularities, regardless of how relatively fleeting those agential interactions in their real-time practices might indeed seem to be in contrast to the perduring institutions they support. Thus, we propose an alternative concept to illustrate how we reside realms of metaphysics in pursuing scientific discovery: epistemic attractors (see Fig. 2). Such a notion of scientific intuition will highlight its core conflict as one consisting of human ingenuity as it is confronted with these “shared examples”, understood as “attractions to assent” \cite{sosa_experimental_2007}. 

We take the notion of an attractor from the mathematical realm of dynamical systems theory, where an attractor is a point or set of points in a multidimensional space which pulls all the activity and change within that space toward itself; once a state within that space reaches an attractor, it remains there forever unless sufficiently perturbed. In our case, we think of an epistemic attractor as a point in epistemic space, a point being a single system of practice, towards which other neighboring systems of practice evolve in the course of history and cultural evolution. As certain activities and operations within a system of practice are altered by their practitioner, either self-consciously or not, they often evolve to a point where their practitioners are satisfied with them or where no further modifications can be made that truly change the material outcomes of those practices. Just as attractors in a dynamical system, these epistemic attractors may be single points (systems of practice) or they may be periodic (historical cycles of systems of practice), or they may be strange attractors where the states of the systems of practice dance around a nebulously defined manifold of activities. Epistemic attractors as a concept have fuzzy boundaries, a Wittgensteinian ``family resemblance'' relationship, and in Wittgensteinian fashion, we will illustrate them ostensively using Hacking's idea of self-vindication in the laboratory sciences as a prime exemplar of the epistemic attractor concept. 

Hacking sets out to explain the apparent stability of the laboratory sciences by showcasing how their varied responses to contradictory evidence have led them to change far more than just auxiliary hypotheses in the course of history. Rather than merely stabilizing the linguistic form of theories in the sciences, these alterations have led to the stabilization of the systems of practice that constitute the laboratory sciences in the first place. Hacking develops his analysis through a rough taxonomy of the practices and objects which have been altered in complex feedback loops throughout the history of science. The taxonomy is broken up into three overlapping categories: \textit{ideas}, \textit{things}, and \textit{marks and the manipulation of marks}. These categories are expanded into their constituent elements as follows:

In \textbf{ideas}, we have

\begin{enumerate}
\item \textit{Questions}: Which phenomena or problems are of interest to the scientists?
\item \textit{Background Knowledge}: Similar to auxiliary hypotheses, essentially non-systematic knowledge or practices which are rarely acknowledged.
\item \textit{Systematic Theory}: High-level theories about some domain.
\item \textit{Topical Hypotheses}: Relatively loose statements which are meant to connect observational terms to theoretical terms; these are related to how theories are verified empirically. 
\item \textit{Modeling of the Apparatus}: Theories concerned with the function and phenomenology of the devices employed in experiments.
\end{enumerate}

In \textbf{things}, we have

\begin{enumerate}
\setcounter{enumi}{5}
\item \textit{Target}: The physical system under study, including its various preparations.
\item \textit{Source of Modification}: Devices for directly interacting with and manipulating the physical system under study. 
\item \textit{Detectors}: Devices which measure the outcomes of interactions with the system under study.
\item \textit{Tools}: A catch-all term for devices which are necessary for experimentation, but which do not fall into the previous categories.
\item \textit{Data Generators}: A broad term that refers to any device that creates data, from a person counting event occurrences to an oscilloscope recording voltages. 
\end{enumerate}

In \textbf{marks and the manipulation of marks}, we have

\begin{enumerate}
\setcounter{enumi}{10}
\item \textit{Data}: Anything produced by a data generator.
\item \textit{Data Assessment}: A type of data processing that involves preliminary procedures such as error estimation and data cleaning. 
\item \textit{Data Reduction}: Any procedure for reducing the dimensionality of data.
\item \textit{Data Analysis}: Activities such as statistical model building, fitting parameters, hypothesis testing, and the like.
\item \textit{Interpretation}: Discussion of what the data and analyses imply for or can be integrated with broader theoretical perspectives. 
\end{enumerate}

Self-vindication refers to fact that the evolution of scientific practices and theories has played out in some ways as a self-fulfilling prophecy: whenever experimentation fails to match a pet theory, investigators may iteratively modify any combination of these aspects of laboratory science not directly related to theory so as to eventually end up with experimental phenomena which do match theory without ever having to give up or alter the theory. Hacking contends that these iterative feedback loops between adjustments of laboratory practices so as to obtain coherent relationships between theory and evidence lead to a stabilization of systems of practice these fields. This is exactly what we mean to capture through the idea of an epistemic attractor, the evolution of a network of investigatory activities toward a fixed point, a system of practice which maintains itself, robust to small perturbations to the individual activities and operations of the system. The system of practice becomes dominated by a unique internal logic that proliferates its coherence and resists inquisition by external systems of practice. 

Hacking's idea of self-vindication is directly related to his concept of self-authentication in a style of reasoning \cite{hacking_language_1982}:

\begin{quote}
    The truth of a sentence (of a kind introduced by a style of reasoning) is what we find out by reasoning using that style. Styles become standards of objectivity because they get at the truth. But a sentence of that kind is a candidate for truth or falsehood only in the context of the style. Thus styles are in a certain sense "self-authenticating." Sentences of the relevant kinds are candidates for truth or for falsehood only when a style of reasoning makes them so. This statement induces an unsettling feeling of circularity.
\end{quote}

An unsettling circularity indeed. The relation between self-vindication and self-authentication is in the polysemous notion of \textit{success}. The criteria for success of an activity involved in the verification of a statement are what make that statement bivalent in the first place. Here we see the relation between Auguste Comte's notion of \textit{positivity} \cite{comte_general_2009} and Dummet's notion of \textit{bivalence} \cite{dummett_what_1996} in terms of styles of reasoning: not only is the positivity of a statement, its possibility of truth-or-falsehood, dependent on a style of reasoning, but also its conditions of verification, its bivalence, are too. 

One consequence of self-vindicatory processes in the laboratory sciences which went relatively unacknowledged until the work of Karen Barad is that even though scientists claim to seek explanations for the same phenomena throughout the course of their work, any adjustments they make to items 6 - 10 in the list above (that is, adjustments to an apparatus) may significantly alter the phenomena under study. In their work on \textit{agential realism} \cite{barad_meeting_2007}, Karen Barad has shown in great detail that in any experimental science, the apparatuses used in experiments play a constitutive role in generating the phenomenon being studied; in other words, the phenomenon does not exist without the apparatus. As illustrated so eloquently in their example of how the concepts of position and momenta of quantum systems co-arise with the devices used to measure them and do not meaningfully exist in the absence of those devices, a minor alteration to the apparatus can radically change the phenomenon in question.

The effects of this continual phenomenon shift through iterative feedback between the various aspects of experimental practices are no more apparent than in the neurosciences. As neuroscientists have reified behavior in laboratory settings to fit the constraints of measurement devices and preconceived ``how-possibly'' explanations, they have altered beyond all recognition the phenomena that they first set out to explain, namely, behavior in the wild. The neurosciences have decayed toward epistemic attractors whose internal logic is capable of generating explanatory fictions fit for publishing papers, but which cannot stand up to testing by any external system of practice that seeks to take ecologically realistic behavior seriously. No neuroscience whose explanations end at the laboratory door can hope to justify its realist pretensions.

\begin{figure}[H]
    \centering
    \includegraphics[scale=0.35]{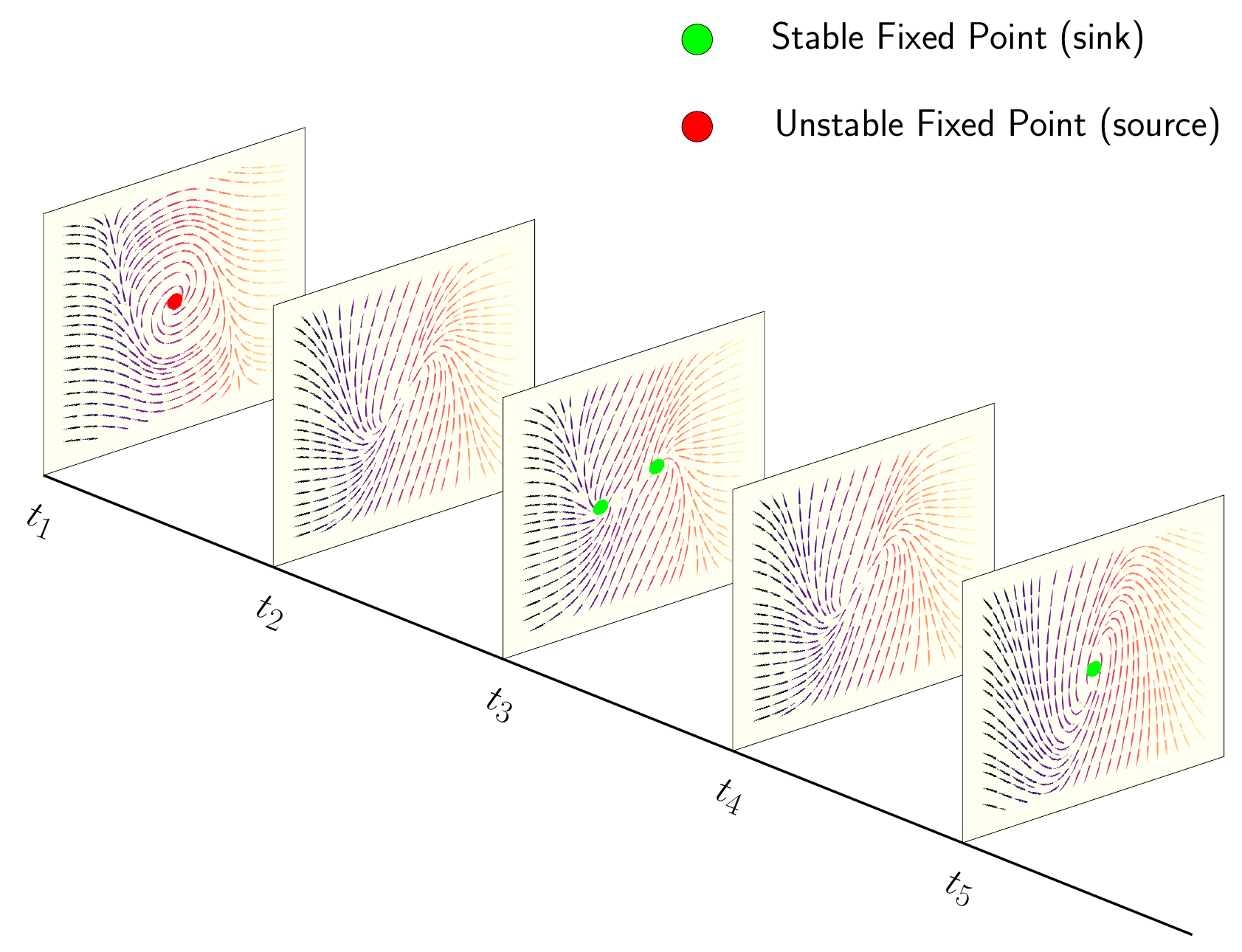}
    \caption{A visual explanation of fixed points in the state space a dynamical system.}
    \label{fig:epiatt}
\end{figure}

\section{Neuroliberalism and the Radically Real}\label{sec4}

\begin{quote}
    "This imperceptible trembling of the finite that makes its limits indeterminate and allows it to blend, to make itself whatever (...) Its beatitude is that of a potentiality that comes only after the act, of matter that does not remain beneath the form, but surrounds it with a halo."\footnote{Agamben, G. (1993: 55)}
\end{quote}

\subsection{The Realist Regime}\label{subsec7}

The realist turn is born as an alternative to the exhausted metaphysic nominalism -fueled by postmodern constructivism and the hegemony of the linguistic turn- that dominated the second half of the 20th century. We understand this antirealist operative consisting in equating ontology with substance-dualism (e.g., identity with representational logic, subject with \textit{res cogitans} and difference with exclusive opposition); we share with Harman \cite{harman2007heidegger} that ontology is neither exhausted and not even fully explored if reduced to the thing-in-itself, as object of a clear and distinct subject. Notwithstanding the rich diversity of the many realisms referred to in passing already and those that will be described in the following sections; regarding their relevance throughout their dialogues with other disciplines -such as genetics \cite{malabou_before_2016}, physics \cite{barad_meeting_2007}, AI \cite{wilson_neural_2016} and mathematics \cite{gabriel_fields_2015}- they coincide in the affirmation of radical immanence and immediate presence of the real; not as an unnamed center-in-itself, remote, unknown and transcendent to the finite world, nor an objectively constituting nucleus external to language, but as a self-acting mediator of meaning and experience. Realist conditions of possibility for accessing knowledge will assume the actuality of object and subject as relational and reciprocal terms of the same reflexive and medial reality in itself and thus a fundamental ontic dual opposition falls apart, namely, extrinsic correlationism between entities (e.g., thing and thought, nature and culture, inner and outer, etc.).

Whenever tales of realism are told, we often find that these attempts focus on minimally viable operational overlaps between theory and practice and, likewise, their convergence is generally limited to a single conceptual domain of narrow epistemological bandwidth. While it shouldn't be of surprise to anyone that incommensurability between disciplines is an inevitable outcome of the particular modes of knowledge productions of modern-, and certainly of those of post-modern origin; once Martin’s (2010) ethnographic argument for the ontological primacy of the sociocultural over the neurobiological is realised, the possibilities of detecting "(...)the real prejudices hidden behind the appearance of objective statements" \cite{latour_why_2004} -that is, throughout the cognitive sciences- shine through.

Consider Fitzgerald and Callard's \cite{callard_rethinking_2015} questions: 

\begin{quote} 
“(...)given that there is the possibility of overlapping interests and objects between these sciences, then how large should that space of overlap be, how should it be populated, what kinds of objects should be located within it, and what should count as a sufficiently ecumenical research programme to address those objects?”
\end{quote}

Whoever has felt estrangement because of surface-level overlapping of cognitive and neuroscientific disciplines, should indeed ponder taking this less as a symbol of their mutual cooperation and much more as a clear symptom of the disunity of science. At this intersection; that is, the one of how meaning is enacted and boundaries emerge from its scaffoldings, a pressing question regarding hegemonic scientific discursive practices is revealed: what is captured and who excluded at the conjuncture of these matters? 

Again, as Kockelman goes \cite{kockelman_epistemic_2020}

\begin{quote}
"In terms of their stakes for human-being, or their role in the production of knowledge, such epistemic dynamics might be usefully compared to a range of other ideas—not just Foucauldian epistemes, Wittgenstein’s language games, and Kuhnian paradigms but also (…) issues pertaining to confirmation holism (via Quine and Duhem), proofs and refutations (via Lakatos), and various kinds of experimental and theoretical regress (via Collins and Kennefick).”
\end{quote}

And, moreover, while we can indeed \cite{halberstam_queer_2011},

\begin{quote}
"(...) think about a kind of theoretical model that flows below the radar, that is assembled from eccentric texts and examples and that refuses to confirm the hierarchies of knowing that maintain the high in high theory"
\end{quote}

there are unfortunately already mechanisms in place which foster the reproduction of a post-colonial "white optics" set of perceptual schemata on agentic matter and new materialist inquiry around an "unnamed center" that comprises implicitly only a specific scope of "certain human-nonhuman assemblages". Neuroscience as a field has been referred to as juvenile yet it was born emerging within a senescent ontology; physicalism, endowed it with the enshrinement of the brain as "a locus of the truth of the self" \cite{feiten_2021, pitts-taylor_plastic_2010} which misconstrues the nature of both experience and that of mind. Merleau-Ponty \cite{merleau-ponty_phenomenology_2006} had already reformulated the Kantian transcendental deduction that the conditions of possibility of subjectivity are coterminous with those of objectivity by stating that the
\begin{quote}
"(...) body is the fabric into which all objects are woven, and it is, at least in relation to the perceived world, the general instrument of my "comprehension".
\end{quote}
Consequently, we need to ask: what kinds of subject are being created in our neuroscientific times? These so-called “mapping practices“ \cite{haraway_simians_1998} are said to materialise bodies together with their particular boundaries as "machine-readable information storage devices" \cite{dijstelbloem_migration_2011}; that is, sources of information and objects of control by means of distribution practices. Foucault had already posited a clear connection between biopolitics and the disciplinary form of capitalism, which inculcates the body to serve its productive purposes. Hence, biopolitics essentially delves into the biological and physical realms, ultimately encompassing a comprehensive politics of the body. However, neoliberalism does no longer ascribe primacy to the "biological, the somatic, the corporal." Instead, it selects the psyche as preferred source of productivity, thereby marking a shift towards psychopolitics (hence, our claim for the advent of neuroscience in times of neoliberalism as the pivoting point for the true psychic turn). Meloni \cite{meloni_how_2014} points out that the central nervous system “(...)has become porous to social and even cultural signals to an unprecedented extent”, given the technological developments of the new brain sciences. Preciado \cite{preciado_yo_2020} further warns us of physicalism having trickled down into psychiatry, too; in the form of a hetero-patriarchal ideology of difference \cite{irigaray1992elemental}; a “social technology” which reduces bodies to their heterosexual reproductive capabilities \cite{butler1993bodies}.

\subsection{The Radically Real}\label{subsec8}

\begin{quote}
"Searching for some motive in the light, squinting through its beams, we say that the material and the social and the symbolic are (...) a productive synthesis. No declared state of Being is independent of experience with what it is like to live. Any process ontology—whether flowing from Bataille or Deleuze or Haraway or Butler—has a pragmatic impulse, seeking after accomplishment in the study of co-development and closer attunement in environmental co-feeling (...) The contours of Things are contours only in a reigning consensus about how Things are (…); the human-independent forces being pursued are found with human hands and felt in human hearts." \footnote{ \cite{gruber_brain_2021}}
\end{quote}

Indeed, western rationalism contains a dualistic violence consisting in rendering a transcendental phase visible that is not \textit{really} inscribed in matter \cite{malabou_before_2016}. This is an old hungover from the century of the lights which transcends physicalism or physicalist reductionism \cite{arendt1998human}. Phenomenology, structuralism, deconstruction and, in particular, postmodernism all have been strains of the continental anti-realist diaspora in rejecting the concern about the Absolute in terms of Reality-in-itself since Kants’ Copernican revolution; while the likes of Hegel, Heidegger, and Derrida have been taken hold of by the re-emergence of Kant’s idealist prohibition (i.e., experience is structured by \textit{a priori} categories and forms of intuition which comprise the necessary and universal basis for all knowledge) as a linkage which Meillasoux \cite{meillassoux_apres_2012} calls “correlationism” or “the idea according to which we only ever have access to the correlation between thinking and being, and never to either term considered apart from the other”.

While Meillasoux recognises modernities mourning of the loss of an external, eternally objective reality, which came with constructivism, he relinquishes transcendence via speculative realism and thus abandons any potential in vindicating the hypostatised “freedom” of sciences, by allowing hegemonic concepts at all levels previously adressed to “fill” and enrapture all "anomalies" \cite{douglas_purity_2005} \footnote{Of which Kockelman \cite{kockelman_meeting_2016} comments: “(…) such strain might be understood as barely evident evidence that some indexical-inferential imaginary, some ground, hermeneutic, or ontology, is “out of touch with,” or “insensitive to,” some world. In particular, such strain consists of all the evidence one might gather (in light of a more inclusive, incipient, or counter, ontology) that the kinds of causal interactions we imagine are incorrect, that the indices we produce and interpret are inadequate, that the individuals we aim our instigations and sensations at are nonexistent, that the distributed agencies we incorporate are incoherent, that our modes of inference are unsound, that our treatment of subjects is unethical (if not diabolical).”} that attempt to “fly below the radar”. Moreover, we argue that constructivism, with all the benefits it might have brought to some social sciences' niches in legitimising them, and to have provided some wiggle-room for nomotetic and ideographic disciplines, it did introduce a bleed right between aspects of difference, thus promoting anti-realism and a technical mistake as far as defining the "value-free" in the constitution of science. The true task of the realist critique lies \textit{in media res}; in midst of that which trough its reification yields epistemic violence towards \textit{what is}.

Notwithstanding whether scientific realism might amount to nothing more than a worthwhile project of utopianism, a re-reading of the cordonning-off of the Absolute in terms of Reality-in-itself  \cite{nishida_ontology_2012}, which does not suffer from the reduction of ontology to the illusory phenomenal but recognizes the co-immanence of ideality and materiality might avoid yielding the rejection of the possibility of Knowledge about an independent world by means of our current infestation of cultural and linguistic marks, so as to attempt to settle what future thought should be cognizant and appropriative about regarding human behavior, as that which ought to be carefully examined in light of what is unamendable, yet neither reified by substantialism nor dissolved in pluralistic relativism. We call for a rejection of the first, for its practical enaction often results in narrow rethorics and frameworks that need to be re-situated as a practice within a social structure with drivers and contexts \cite{choudhury_critical_2009} that surpass the "contempt for anything that limits the kind of commensurability that our markets and systems of governance demand" \cite{martin_potentiality_2013}; this is, a general condescension, at the expense of which science has ceased to be understood as a cultural activity itself. We warn against the second, for its stale ethic guides interaction between the cognitive disciplines based on a concern of their "before" as the utmost limiting factor for the ontic update necessary in a collaborative turn, which should not capitulate in face of pre-existing separateness.

Perhaps one could conceive a point of entry by considering research as an assemblage \cite{fox_sexuality-assemblage_2013, coleman_deleuze_2013, masny_rhizoanalytic_2013} which comprises the abstractions that get caught up in social inquiry, including the events that are studied, the tools, models and precepts of research, and the researchers. Roepstorff \cite{roepstorff_enculturing_2010}, for instance, invites to re-think forms of social interaction as ‘patterned practices’, in the context of regarding experiments as a form of \textit{avant-garde performance}; an aesthetic as much as a methodological rubric\footnote{Just as Cazeaux \cite{cazeaux_metaphor_2007} wants to cut nature at is joints after realising that "(...) the categories I need to make my experience intelligible are also those which I find apply to the bits and pieces of the world".}. And yet, while we \cite{ladyman_every_2007}

\begin{quote}
"(…) may not be able to think about structure without hypostatizing individuals as the bearers of structure, (…) it does not follow that the latter are ontologically fundamental",
\end{quote}

the subject requires true life and not just experience to “shine through” and \textit{defract} (i.e., orient) our scientific aesthetics and rituals; even if only by means of a mercurial state of uncertainty, there’s an epistemic requirement for cognitive neuroscience to remain close to this kind of realism by actively attempting to articulate a non-pluralistic pragmatism akin and indeed familiar to contemporary ethnographic stances, where non-pluralism stands -in short- against binary reductionisms (which pluralism tolerates); sexual difference, for instance, is nowadays frequently reduced either to an essentialist biologicism or to a heternormative sociologism that hides the true nature of the hegemonic system \cite{jeffreys_unpacking_2003, raymond_passion_2001}. 

What is radical about realism, which cognitive neuroscience should learn from, then? The unamendable resistance of reality lies in its immediacy being that which is irreducible to mere discourses texts, socio-economic relations or cultural constructions, which amount to abstract substantialisms and sociolinguistic constructivism, respectively. Rendering it intelligible is to effectively perform an epistemic unblocking -the task at hand- so as to deliver power to which only ever receives an opportunity to speak up but is also constrained to only that. As Roland-Rodríguez \cite{roland-rodriguez_can_2020} warns:

\begin{quote}
 “Current debates are not calling for a mere revisitation of the Kantian transcendental, but for a resolution to the conflict that this has with the ontological causes of the individual organism in relation to its consciousness. In other words, just as an experience [vécu] came to identify itself as from or in a body [vivant], so too does metaphysical wonder come to ask itself whether it is also from or in a body, entangled with our genealogy of life (…) this time, the transcendental will not be the key to thinking or solving the metaphysics of life from some place outside, but rather found within life.”
\end{quote}

Barad \cite{barad_meeting_2007} has asserted an interesting midway solution for the problem that realism faces since the emergence of social construction: as opposed to Descartes' division of mind and the unanimated, she asserts the existence of an entanglement of matter and meaning:

\begin{quote}
"Realism...is not about representations of an independent reality but about the real consequences, interventions, creative possibilities, and responsibilities of intra-acting within and as part of the world".
\end{quote}

It shall remain of critical importance to observe such an interstitial field as cognitive neuroscience as a discernible one, and not jump to the hasty conclusion of it being composed only of "non-personal, ahuman forces, flows, tendencies, and trajectories" \cite{bennett_vibrant_2010}, as the antirealist twilight of postmodernity was fortunately brought about by unamendability \cite{ferraris_introduction_2015}:

\begin{quote}
"(...) something, or rather, much more than we are willing to admit, is not constructed - and this is a wonderful thing (...)“.
\end{quote}

\begin{figure}[H]
    \centering
    \includegraphics[scale=0.2]{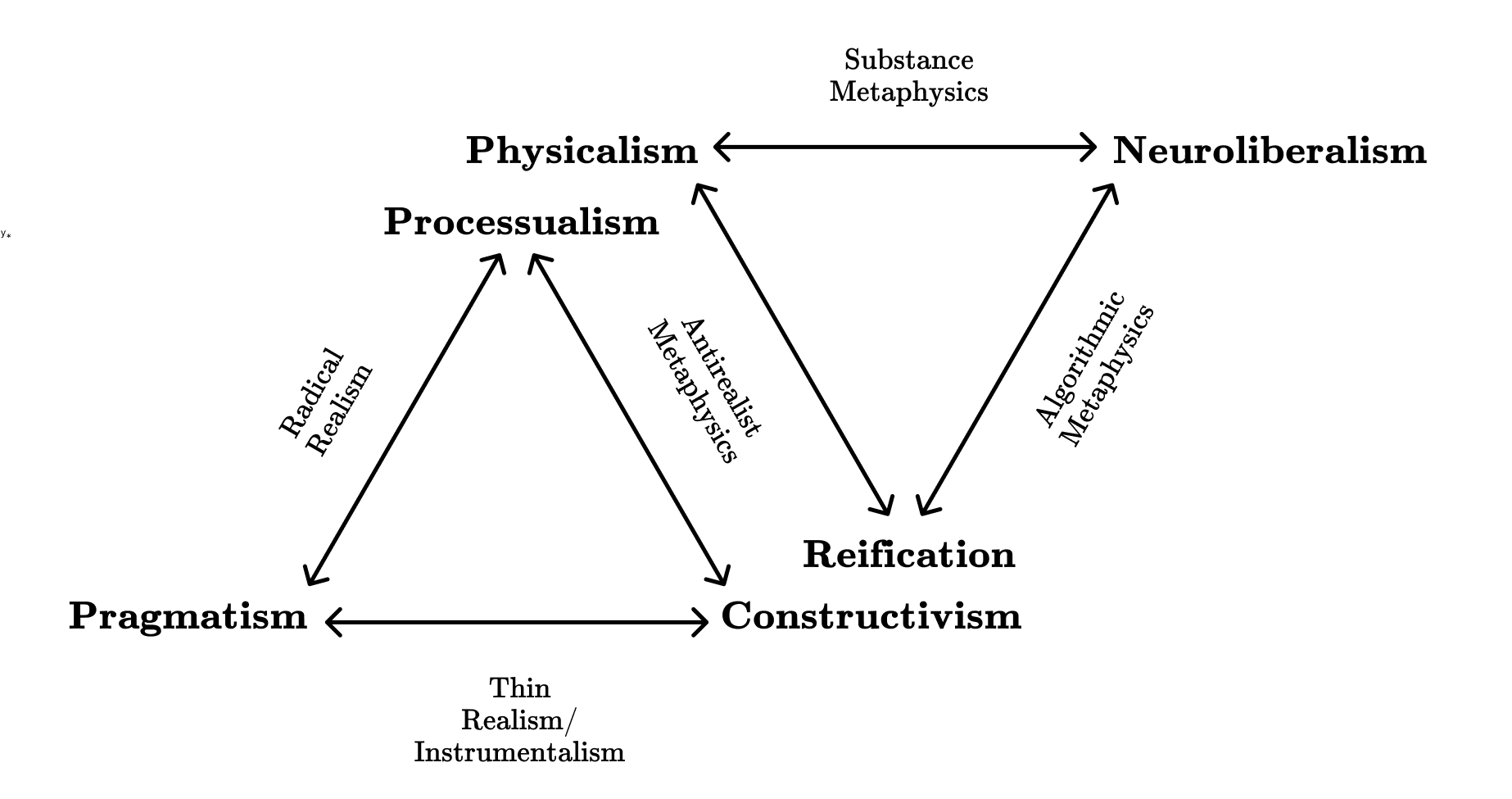}
    \caption{Edge are dynamical spaces wherein epistemic attractors are inscribed. Corners are disciplines, paradigms, and mechanisms which serve as the foundations for the different realisms discussed in the text.}
\end{figure}

\section{The Mind-Independence of Success}\label{sec5}

\subsection{Hasok Chang's Pragmatist Realism}

Hasok Chang, recognizing the difficulties plaguing standard success arguments for scientific realism, has attempted to remedy those difficulties using a pragmatist notion of truth that leads him to frame realism as an inherently pluralist doctrine. Chang’s pragmatic realism emerges out of concerns regarding correspondence theories of truth and their relation to standard scientific realism. Chang asks us to radically alter how we think about science; instead of thinking of science in terms of disembodied theories (networks of propositions), pragmatic realism suggests that we should subordinate \textit{knowing-that} to \textit{knowing-how} and think of science in terms of the actions of its practitioners \cite{chang_realism_2022}.

One might summarize Chang’s position by one of his stated goals: ``killing correspondence'' \cite{chang_realism_2022}. For Chang, the notion of correspondence between statements in our scientific theories and an “external world” which those statements represent is a metaphor arising in fictional narratives we tell each other, exercises of the imagination. We observe actual human representational practices, such as when a child sketches an image of a house, and imagine that our theories are akin to these drawings: there must be a house (the noumena, external world, things-in-themselves, etc.) which the drawings (theories) represent. 

Chang seeks to replace the notion of correspondence with that of operational coherence. When philosophers talk about coherence, they are usually referring to logical relations among sets of propositions (e.g. the set is free of contradictions). In order to distinguish his activity based concept of coherence from the logical one, Chang refers to logical notions of coherence as \textit{consistency} and his activity based notion of coherence as \textit{operational coherence} \cite{chang_realism_2022}. For Chang, operational coherence is roughly “a harmonious fitting-together of elements and aspects of an activity, which is conducive to the successful achievement of the aims of that activity” \cite{chang_realism_2022}. With this notion of operational coherence at hand, we can return to Chang's tripartite activity-based analysis of science to more formally define an epistemic activity and system of practice:

\begin{quote}
    \textbf{Epistemic activity}: a \textit{coherent} set of mental or physical operations that are intended to contribute to the production or improvement of knowledge in a particular way, in accordance with some discernible rules. 
\end{quote}

\begin{quote}
    \textbf{System of practice}: a coherent and interacting set of epistemic activities performed with a view to achieve certain overall aims.
\end{quote}

Chang more formally defines operational coherence by the following statement \cite{chang_realism_2022}:

\begin{quote}
    An activity is \textbf{operationally coherent} if and only if there is a harmonious relationship among the operations that constitute the activity; the concrete realization of a coherent activity is successful, \textit{ceteris paribus}; the latter condition serves as an indirect criterion for the judgement of coherence.
\end{quote}

What the last two clauses could mean is that we may initially discover the way to successfully execute some action by accident, or by some other process of variation and selection, but once we know how to achieve the activity in some context, we can say under what conditions it will be successful in that context. If the context changes, the conditions for success may change as well.

But the notion of success that Chang has in mind is different than mere accidental discovery. Chang is using ``success'' to refer to events or phenomena which are aligned with our pre-defined goals. We define the conditions under which some epistemic practice will be successful and then check whether it is. There is a disturbing similarity here with Hacking's processes of self-authentication and self-vindication as well as between Hackings's styles of thinking and Chang's systems of practice. Indeed, Chang's idea of ``truth'' is really just a refined version of Hacking's notion of ``positivity'' \cite{chang_realism_2022}:

\begin{quote}
    A statement is true to the extent that there are operationally coherent activities that can be performed by relying on its content.
\end{quote}

Which leads immediately to Chang's conception of what makes something real \cite{chang_realism_2022}:

\begin{quote}
    An entity is real to the extent that there are operationally coherent activities that can be performed by relying significantly on its existence and its properties. (And \textit{a reality} is a real entity). 
\end{quote}

We will begin the next section by suggesting possible solutions to the issues we have highlighted in Hasok Chang's pragmatic realism. Though he tries to detach from the idea that theories represent reality, even explicitly acknowledging representationalism's fatal flaws, Chang fails to reach escape velocity. He still speaks of the referents of words, the entities in our ontologies. Though many realists may simply want a theory that tells them what should and should not be considered real, we are not satisfied with such theories. We want a notion of empirical success that is independent of our human aims and a corresponding explanation of why some actions are successful and others not. 

Note that amidst all this, Chang is focused on a picture of reality in terms of \textit{things} rather than \textit{processes} or \textit{change}. The assumptions of substance metaphysics and its focus on stable objects continue to haunt even the seemingly most radical perspectives on realism available today.

\subsection{Desiderata on Relata}\label{subsec5}


Words do not have referents, they do not represent entities. What matters is not so much the words we use, but what they make us do. To the extent that language has meanings at all, such meaning is in virtue of what actions it affords to us (and yes, \textit{thought is a kind of action}). If we hope to have a realism worthy of the term, a neuroscience with relevance to the explanation and understanding of natural behavior in its ecological context, we must resist the acceptance of self-authenticating process in science. Consider what Hacking says about self-authentication \cite{hacking_language_1982}:

\begin{quote}
    The apparent circularity in the self-authenticating styles is to be welcomed. It helps explain why, although styles may evolve or be abandoned, they are curiously immune to anything akin to refutation. There is no higher standard to which they directly answer.
\end{quote}

Although styles of reasoning or systems of practice (more generally, sense-making) may not recognize a higher standard to which they must answer, we believe that there is indeed a higher standard to which they \textit{should} answer. If we consider what Hinton et al. \cite{hinton_2015} insist upon, namely

\begin{quote}
“(…) an unavoidably enfolded methodological gesture – that is, when we inquire we co-constitute – the very stuff of our analysis is a dynamic materialising of both the object and the subject of our epistemic interventions." 
\end{quote}

Then we must ask, what makes sense-making possible? What determines which activities can be successful and how their success comes about? Using Hasok Chang's language, what dictates whether the relationships among the operations that constitute an activity are harmonious? To address this, we turn to inspirations from New Realism.

Though a number of scholars have joined in the emerging New Realist movement, we take Ferraris’ position to be characteristic of New Realist views \cite{ferraris_introduction_2015}. Ferraris, raised in the postmodernist tradition, begins his negative project in doubt about his mentors’ social constructivism. In one sense, Ferraris picks up where Chang leaves off. Chang’s pragmatic realism makes reference to \textit{successful actions} in defining the real, but neglects to give an account of what allows some actions to be successful and others not, or in other words, what non-human processes dictate whether an action will be successful or not. This is Ferraris’ concept of \textit{unamendability}: the real is independent of our knowledge of it and may often resist our efforts to intervene on it when those interventions are not guided by that which makes successful action possible \cite{ferraris_introduction_2015}.

But here we are in need of a notion of success that is also independent of the narratives which we use to frame our goals. A radical realism, characterized by such a notion of success in science, would not merely be another inert theory in the metaphysics of science; it would be capable of pragmatically shaping scientific activities into systems of practice not bound by a purely internal coherence incapable of bearing external scrutiny. We require a concept of empirical success not based in predefined criteria which lend themselves to self-authenticating and self-vindicating dynamics in epistemic space. The inevitable question comes: where are we to find such a concept?  

A hint at a possible solution comes to us from Mark Bickhard's interactivist theory of cognition, which draws heavily upon process metaphysics and Gibsonian psychology \cite{bickhard_interactivist_2009, gibson_senses_1983}. According to Bickhard, all cognition emerge as a result of processes of variation and selection whereby the organism discovers lawful regularities between the structure of its environment and its own activities in the environment \cite{bickhard_variations_2003}. The selection of particular (initially random) behavioral variations occurs due to anticipatory processes within the organism: violations of implicit predictions of future states of the organism induce learning processes which eventually cement context appropriate behavioral trajectories. For Bickhard, anticipatory dynamics within the organism emerge due to various thermodynamic constraints, and it is the historical selection of behavioral patterns according to anticipatory dynamics which implicitly defines a notion of successful behavior. If we can find a way to tell a story scaffolding the variation and selection of low-level motor movements into high-level scientific practices, then we can draw from the implicit human-independent notions of successful behavior to define success of scientific practices in a way that is agnostic to how we frame our scientific goals. 

More work is needed to flesh out a theory of empirical success based on interactivist conceptions of variation and selective retention. What is clear though is that neuroscience (and the sciences in general) will benefit from shedding the baggage of traditional substance metaphysics and representational approaches to realism.

\bibliographystyle{unsrt}
\bibliography{manuscript}

\end{document}